\documentclass[preprintnumbers,amsmath,amssymb,twocolumn]{revtex4}
\usepackage{eso-pic,calc}
\usepackage{graphicx}
\usepackage{amsthm}
\usepackage{epsfig}
\usepackage{bm}
\usepackage{color}
\usepackage[colorlinks,linktocpage,linkcolor=black]{hyperref}
\usepackage{mathrsfs}

\def\beqr{\begin{eqnarray}}
\def\eqnr{\end{eqnarray}}
\def\beq{\begin{equation}}
\def\bc{\begin{center}}
\def\ec{\end{center}}
\def\eqn{\end{equation}}

\def\rmp#1#2#3{{ Rev. Mod. Phys.} {\bf #1}, #2 (#3)}
\def\prl#1#2#3{{ Phys. Rev. Lett.} {\bf #1}, #2 (#3)}

\def\prb#1#2#3{Phys. Rev. B {\bf #1}, #2 (#3)}

\def\pre#1#2#3{Phys. Rev. E {\bf #1}, #2 (#3)}

\def\epl#1#2#3{{ Euro. Phys. Lett.} {\bf #1}, #2 (#3)}
\def\epjb#1#2#3{{ Euro. Phys. J. B} {\bf #1}, #2 (#3)}

\def\physa#1#2#3{Physica A {\bf #1}, #2 (#3)}

\def\natcom#1#2#3{Nat. Commun. {\bf #1}, #2 (#3)}
\def\natphys#1#2#3{Nat. Phys. {\bf #1}, #2 (#3)}
\def\sc#1#2#3{Science {\bf #1}, #2 (#3)}

\def\etc{ etc.}
\def\ie{i.e.,~}
\def\etl{$et~al.$~}
\def\la{\langle}
\def\ra{\rangle}

\begin{document}

\title{Scaling of Extreme Events in 2d BTW Sandpile}

\author{Abdul Quadir}
\author{Haider Hasan Jafri}
\affiliation{Department of Physics, Aligarh Muslim University, Aligarh, 202 002, India}

\begin{abstract}
We study extreme events in a finite-size 2D Abelian sandpile model, specifically focusing on avalanche area and size. Employing the approach of Block Maxima, the study numerically reveals that the rescaled distributions for the largest avalanche size and area converge into the Gumbel and  Weibull family of Generalized Extreme Value (GEV) distributions respectively.  Numerically, we propose scaling functions for extreme avalanche activities that connect the activities on different length scales. With the help of data collapse, we estimate the precise values of these scaling exponents. The scaling function provides an understanding of the intricate dynamics within the sandpile model, shedding light on the relationship between system size and extreme event characteristics. The findings presented in this paper give valuable insights into the extreme behaviour of the Abelian sandpile model and offer a framework to understand the statistical properties of extreme events in complex systems.
\end{abstract}

\maketitle

\section{Introduction}

A complex system refers to a system composed of numerous interconnected components or elements. The interactions among these components give rise to emergent behaviour that is often difficult to predict from the properties of the individual parts. The widespread existence of complex networks in both natural and societal contexts, such as interlinked biological and chemical systems, neural networks, the internet, the WWW, and social networks, indicate that complexity is pervasive~\cite{Dorogovtsev_2003, Caldarelli_2007, Barrat_2008, Barabasi_2016, Albert_2002}. Because of the fluctuating nature of these systems, with their inherent intricacies and differences in properties, they have been a subject of continuing interest in the past few decades. The dynamics of complex systems may give rise to large fluctuations in a relevant variable, resulting in extreme events that may be defined as events exceeding a predefined large threshold. These events may have cascading effects throughout the system and may deviate significantly from a system's average (or usual) patterns. They are often rare but can have profound and disproportionate impacts. Extreme events occur in numerous scenarios, namely, the breakdown of a mechanical structure, an earthquake, flooding or crashes in financial markets \cite{Montroll_1974}. Other instances where these events have been reported are the systems that exhibit self-organized critical phenomena~\cite{BTW, Tang_1988, Bak_1996, Dhar_1989, Dhar_1990, Dhar_2006}, spontaneous brain activity, fracture~\cite{Sahini_1993}, portfolio management, Darwinian evolution of fitter proteins~\cite{Weinreich_2006}, fickle stock exchange~\cite{Goncalvesab_2011}, acute scenarios in capricious weather~\cite{Cai_2014, Katz_1992}, seismicity risk evolution and other geophysical processes.

In the complex systems theory framework, it is important to understand and classify the possible underlying mechanisms responsible for huge fluctuations. Further, one can also derive statistical properties of the underlying statistical distributions. The classical theory that offers a systematic approach to understanding the statistical properties of rare (or extreme) events is called as the extreme value theory (EVT). This allows for a deeper exploration of tail behaviour beyond the scope of conventional statistical methods. EVT has been studied in engineering~\cite{Gumbel_1958, Weibull_1951}, finance~\cite{Embrecht_1997}, hydrology~\cite{Katz_2002}, meteorology~\cite{Storch_2002}, Pinning-depinning dynamics~\cite{Yan_2024} and natural sciences~\cite{Katz_2002, Storch_2002, Yan_2024, Gutenberg_1944, Bouchaud_1997}, to name a few. Traditionally, the EVT is categorized into three extreme value statistics limit distributions: Fr\'echet, Fisher-Tippett-Gumbel (FTG), and  Weibull. In the literature, numerous studies have focused on the studies of finite size effects using the renormalization-group (RG) theory and proved that RG applies regardless of whether these variables are independent or correlated~\cite{Coles_2001, Haan_2006, Galambos_1978, Albeverio_2006, Calvo_2012, Fortin_2015, Bramwell_2009, Castillo_2005, Clusel_2008, Bramwell_2000, Bertin_2005, Ghil_2011, Schehr_2006}. In 2d site percolation problem, since the cluster size in sub-critical site percolation follows $p(x) \sim x^{-1}e^{x/x_c}$, with $x_c$ as cutoff~\cite{Stauffer_1994}, the largest cluster size follows FTG distribution~\cite{Gyorgyi_2008, Gyorgyi_2010}. The EVT in $1/f^{\alpha}$ suggests FTG distribution for $0 \leq \alpha < 1$ whereas it follows a nontrivial distribution for $\alpha>1$~\cite{Antal_2001, Antal_2009}. In the case of the fitness model for the scale-free networks having nodes with homogeneous fitness, the degree distribution converges to the Gumbel distribution. However, the distribution converges to the Fr\'echet distribution in the case of nodes having heterogeneous fitness~\cite{Moreira_2002}. 

The presence of many degrees of freedom in the complex system leads to the critical phenomenon and the emergence of long-range correlation. Such long-ranged phenomena that exhibit scale invariance in the absence of an external tuning parameter is known as  ``Self-organized criticality (SOC)''. The SOC systems are characterized by their tendency to evolve towards a critical state~\cite{BTW}. In this situation, the observable quantities display power-law distributions. P. Bak, C. Tang, and K. Wiesenfeld's (BTW) hypothesis elucidates the emergence of scaling in a slowly driven non-equilibrium system found in nature~\cite{BTW, Bak_1996, Tang_1988}.  The SOC systems are an important class of complex systems that are capable of generating extreme events. Understanding extreme events in the SOC systems is important because it provides insight into the robustness and vulnerability of these systems, helps assess the potential risks associated with rare events, and further contributes to the understanding of self-organizing systems. Additionally, the study of extreme events in the SOC systems has implications for risk management, disaster preparedness, and the resilience of systems in the face of unpredictable and impactful events. The Abelian BTW sandpile with infinite system size is critical, and the behaviour of extreme activities can be easily derived~\cite{Katz_1992}. However, in real situations, the systems have a finite size, which may induce correlations, cutoffs and system size effects in the events~\cite{Dhar_1989, Dhar_2006, Dhar_1990, Yadav_2022}. Garber~\etl suggested that the extreme avalanche size probability distributions are affected by finite-size (FS) effects~\cite{Garber_2009} and the nature of the distribution crucially depends on the range of observation or the block size.

In this work, we show that the extreme value distribution (EVD), associated with extreme activity, scales with the system size. We highlight that the probability and parameters may vary with the system size and belong to the same class of generalized extreme value distributions (GEVD).  We demonstrate a simple scaling analysis that can capture this characteristic. The method depends on identifying the characteristics of scaling functions for GEVD.  The rest of the article is organized as follows. In Sec.~\ref{Sec:GEVT}, we describe the generalised extreme value distribution. In Sec.~\ref{Sec:BTW}, we study the finite size Abelian BTW sandpile model as a particular model and characterize the distribution as {\it Gumbel-Fr\'echet-Weibull} universality classes for the magnitude events. Further, we present a proposed universal scaling mechanism for extreme events in Sec.~\ref{Sec:USF}. We present our findings and the proposed FS scaling for extreme avalanche activities. Finally, in Sec.~\ref{Sec:Conclusion}, we draw our conclusions.

\section{Generalized Extreme Value Theory}~\label{Sec:GEVT}
To understand the statistics of extreme events, we consider the distribution of the maxima of the observable quantity. The maxima are obtained by dividing the dataset into intervals of fixed length. Then the maximum value from each block is considered. Let $x_i \in max\{X_1, X_2,.... \}$ be the maxima of the independent and identically distributed random observations in each block $X_i$. It was shown that the distributions of maxima $X_i$ may follow a single peak probability distribution~\cite{Gumbel_1958}. The cumulative distribution function (CDF) for the maxima $x_i$ results in the GEV distribution, which is given as~\cite{Coles_2001, Galambos_1978, Haan_2006, Albeverio_2006}
\beq
\mathcal{F}(x,\mu,\beta,\xi) = \exp \Big\{ - \Big[ 1 + \xi \Big( \dfrac{x - \mu}{\beta} \Big) \Big]^{-1/\xi} \Big\} ~\label{Eq:GEVD1}
\eqn
where $\mu,~\beta$ and $\xi$ are location (or mode), scale, and shape parameters, respectively, having bounds $ \mu,~\xi \in \mathbb{R} $ and $ \beta \in \mathbb{R} \mid \beta > 0 $. Depending upon the value of shape parameter Eq.~\eqref{Eq:GEVD1} can be categorized into three universality classes~\cite{Gumbel_1958}. For $\xi > 0$, Eq.~\eqref{Eq:GEVD1} reduces to Fr\'echet class where the parent distribution decays as a power law. The case $\xi<0$ describes the  Weibull class for which the parent distribution decays faster than the power law. For these cases, the corresponding probability distribution function (PDF) is given by
\beqr
f(x, \mu, \beta, \xi) = & \dfrac{1}{\beta} \left( 1 + \xi \dfrac{x- \mu}{\beta}  \right)^{- (\xi + 1)/\xi} \nonumber \\
& \exp \Big\{ - \Big[ 1 + \xi \Big( \dfrac{x-\mu}{\beta} \Big) \Big]^{-1/\xi} \Big\} ~\label{Eq:PDF-GEV}
\eqnr

In the limit, $\xi \rightarrow 0$, Eq.~\eqref{Eq:GEVD1} takes the form

\beq
\mathcal{F}(x, \mu, \beta) = \exp \Big\{ -  \exp \Big(  \dfrac{x-  \mu}{\beta} \Big) \Big\}~\label{Eq:Gumbel-CDF}
\eqn

which describes the {\it Gumbel} class, and the corresponding PDF is given by
\beq
f(x, \mu, \beta) = \dfrac{1}{\beta} \exp\Bigg\{ -\exp\left( - \dfrac{x - \mu}{\beta} \right) - \left( \dfrac{x - \mu}{\beta} \right) \Bigg\}~\label{Eq:Gumbel-PDF}
\eqn

\section{Model}~\label{Sec:BTW}

In this section, we consider a 2d Abelian BTW sandpile model~\cite{BTW} to formulate the EVT. We consider a square lattice of size $N = L^2$, where $L$ is the linear extent. Each site of the lattice is assigned a variable $E_i$ that may describe the discrete height or slope. This dynamical variable could be any physical quantity such as grain density, stress, height, or energy \etc~ The threshold of this variable for each site is taken to be $E_i^{th}$. To drive the system, we update a randomly chosen site with $E_i \rightarrow E_i + 1$. A site $i$ may become unstable if $E_{i} > E_i^{th}$ and hence follows the relaxation rule given by, 
\beqr 
E_i &\longrightarrow E_i - E_i^{th} \nonumber \\
E_j &\longrightarrow E_j + \delta E, \nonumber 
\eqnr 
where $i$ is the unstable site having $j$ nearest neighbours. We consider the equal threshold for all the lattice sites \ie $E_{i}^{th}= E^{th}=4$ and $\delta E =1$ as described in the original Abelian BTW sandpile \cite{BTW}. The grains can leave the system at the boundaries, indicating that the system is dissipative. This rule is repetitively applied until all the sites become stable and the avalanche (or activity) stops. To ensure the relaxation of the time scales, new driving occurs only after an ongoing avalanche is over. In the present study, the observables of interest are the avalanche size $(s)$ and area $(a)$. Therefore, we parameterize the avalanches by variables $x \in \{s, a \}$: namely, the avalanche size $(s)$ defined as the total number of topplings and the avalanche area $(a)$ that gives the total area affected by the avalanche.

In situations where the system size of a sandpile model is finite, it was shown that the observable $x$ follows the power law with a cutoff
\beq 
P(x,x_c) = \begin{cases}
            Ax_c^{-\theta} x^{-\tau_x},  &   \text{\ \ for \ \ } x \ll x_c \\
            \text{rapid change},         &   \text{\ \ for \ \ } x \approx x_c      
            \end{cases}~\label{Eq:PDF1}
\eqn
where $\tau_x$ is the critical exponents, $\theta$ is the scaling exponent, and the upper cutoff appears in the power-law is approximated by  $x_c \sim N^{D_x}$ where $D_x$ is the cutoff exponents and it is found that $ \tau_x + \theta = -1.22$~\cite{Yadav_2022}. However, the BTW model with infinite system size exhibits the power law with no cutoff, which implies ``no largest events" and ``undefined mean". The observation is different in the case of the finite system size~\cite{Garber_2009} and we shall explore the effect of system size on extreme events in the case of the SOC model.  In this case, the rare (or extreme) events are the largest activity in a particular block (or range) of observation. The finite size of the system introduces an upper limit of the avalanche activity, and slow drive may introduce intermittency in the avalanches of any magnitude~\cite{Christensen_2005}.

\section{Extreme Activities in the BTW Model and System Size Scaling}~\label{Sec:USF}

\begin{figure*}[htb]
    \centering
    \includegraphics[width=170mm, height=120mm,angle=0]{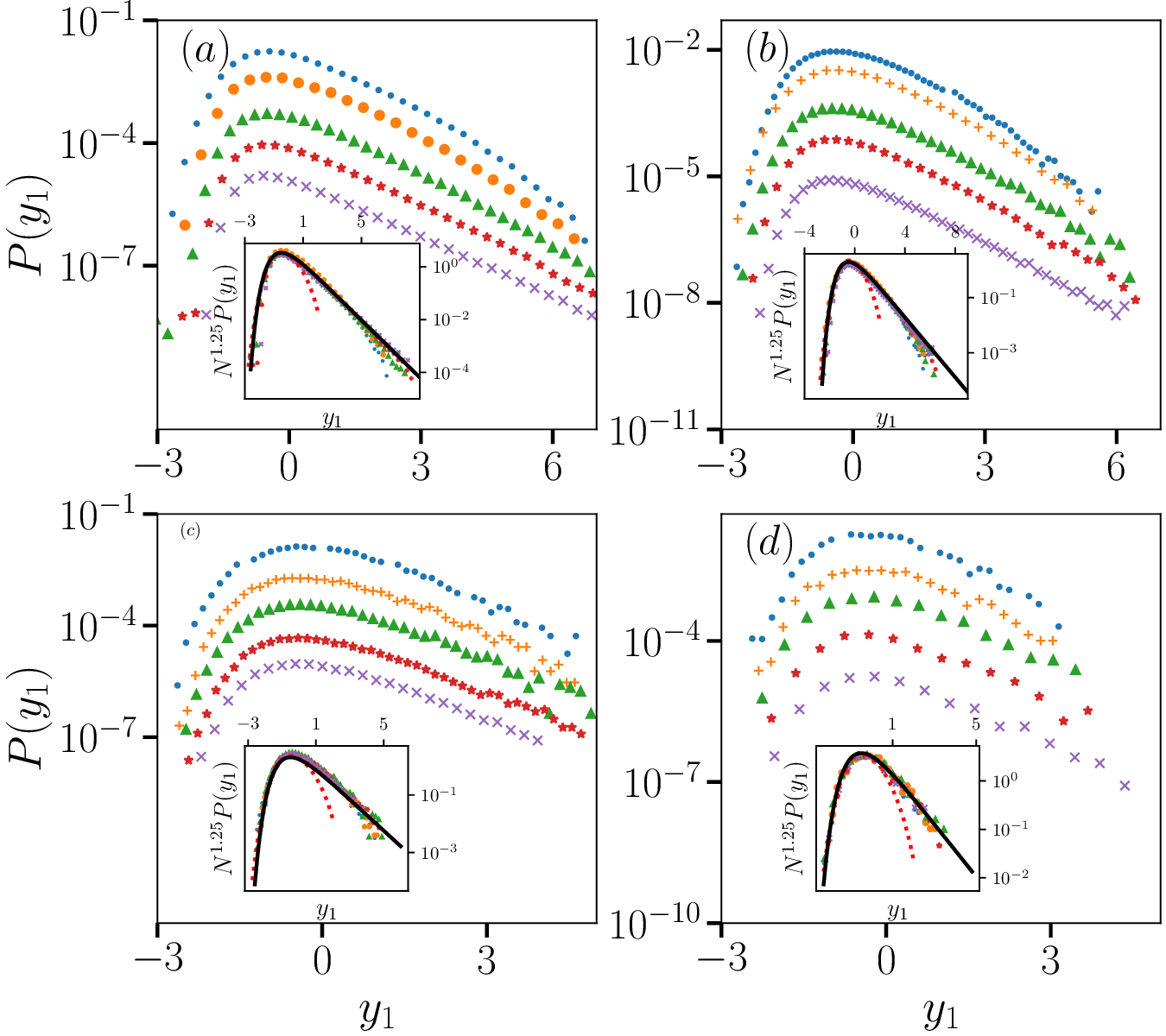} 
    \caption{Probability distribution of the GEV for extreme avalanche size corresponding to different values of the system size $L^{2i}$ using blue dots $ (i=3$), yellow points ($i=4$), green points ($i=5$), red points  ($i=6$), and purple points ($i=7$). These results are plotted for different block sizes equal to (a) $10^2$, (b) $10^3$, (c) $10^4$, and (d) $10^5$. In the inset, we plot the data collapse curve by plotting $N^{\gamma_1}P(y_1)$ versus $y_1$ for different block sizes. The black curve in the inset is the plot for the Gumbel  family and the red dashed line shows the curve from the Weibull family.}
    ~\label{fig-y1}
\end{figure*}

In this section, we shall study the effect of system size on the statistics of extreme events. To generate these events, we consider a time series of the avalanche activity $z_i$, $i=1,\dots, T$, where $T$ is the length of the time series. The series is then divided into $M$ blocks ($M \in \mathbb{N}$) such that the avalanche activity is given by  $\{z_1, z_2,.....,z_M \}$. Extreme events are the maxima over the time series defined as $z^{max} = max\{z_1, z_2,.....,z_M\}$. The set $x_k = \{ z_i^{max},~\forall~i \}$ may follow a single peak extreme value distribution as discussed earlier (Sec.~\ref{Sec:BTW}). Consider the distribution followed by $x_k$ is given by,
\beq 
\mathcal{G}(x_k; N) = N^{-\gamma} \dfrac{1}{\sigma_k} \mathscr{F}(x_k,\la x_k \ra, \sigma_k)
\eqn 
where $N$ is the system size and $\mathscr{F}$ is the GEV function given by Eq.~\eqref{Eq:GEVD1}, $ \la x_k \ra$ and $\sigma_k$ is the mean and the standard deviation of the associated extreme activities. We include a multiplicative pre-factor $N^{-\gamma_k}$ to account for the finite system size effect with $\gamma_k$ as the scaling exponent. Further, we normalize the extreme variables as $y_k = \dfrac{ x_k - \la x_k \ra }{\sigma_k}$ to obtain
\beq 
P(y_k;N) = N^{-\gamma_k} G(y_k)~\label{Eq:Scale}
\eqn
where $G(y_k)$ is the scaling function which helps to isolate the family of GEV. This proposal is fairly general and may be applied to other nontrivial SOC systems that show extreme activities.

\begin{figure}[htp]
    \centering    
    \includegraphics [width=60mm,height=50mm,angle=0]{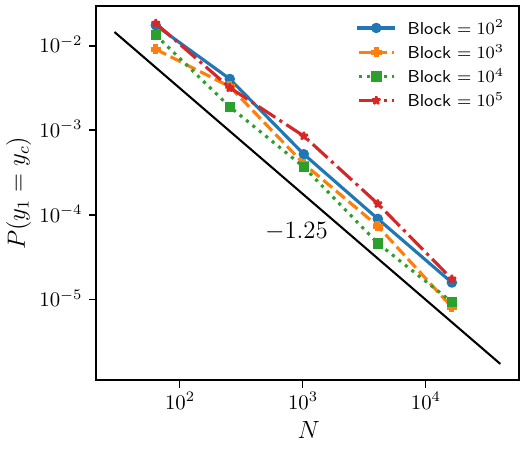}
    \caption{ $P(y_1=y_c)~vs~N$ along with the fitted (black solid) curve $\sim N^{-1.25}$ for the rescaled extreme avalanche size $y_1$}
     \label{Fig-sc}        
 \end{figure}
 
In simulations, we collect the avalanches after discarding transients. The initial configuration is chosen randomly but close to the threshold value. Such a choice is useful in reducing transient time. In the long-time series of avalanches, we implement block maxima methods to identify the largest activity. We keep the total number of largest activities (events) fixed to observe a precise system size dependence. The value of these events should be large enough to minimize the statistical error. To compute the scaling function and identify the family class of the GEVD, we determine the exponents $\gamma_k$ and $\xi$ (cf. Eq.~\eqref{Eq:PDF-GEV} and ~\eqref{Eq:Scale}).  In the following, we shall calculate the value of $\gamma_k$ with the help of the probability distribution function of the system for varying system sizes.

We consider the 2d Abelian sandpile model described in Sec.~\ref{Sec:BTW} and we collect $10^8$ maximum values of the avalanche activities for the avalanche size ($x_1=s$) and the avalanche area ($x_2=a$) over different block sizes. Such simulation is performed for different system sizes $N=L^2,~L=2^i$, for a finite range of $i$ values. In the 2d-BTW sandpile model, the amplitude of extreme avalanche activities may increase as the system size increases. This implies that the mean and standard deviation explicitly depends on the system size, namely, $\la x \ra  \equiv \la x(N) \ra $ and $ \sigma \equiv \sigma(N) $. Further, we observe a systematic variation of $\la x (N) \ra$ and $\sigma(N)$ with system size $N$ and it follows $\la x (N) \ra \sim N^{\lambda_1}$ and $\sigma (N) \sim N^{\lambda_2}$ with $\lambda_i > 0$ (not shown here). Thus, we normalize the avalanche activities.

To explore the effect of system size, we plot the probability distribution of the maxima of the normalized variable $y_1$ corresponding to the avalanche size ($x_1$) given by 
\beq
y_1=\frac{x_1-\la x_{1} \ra}{\sigma_1}
\eqn
where $\la x_{1} \ra$ and $\sigma_1$ are the mean and standard deviations for the variable $x_1$.
The probability distribution of maxima ($P(y_1)$) for various system sizes $N=L^2,~L=2^i$, $i=3,4,5,6,7$ is shown in Figs.~\ref{fig-y1}. As shown, we plot the probability distribution by dividing the variable $y_1$ into various blocks ($M$). The probability $P(y_1)$ for the block sizes $M=10^2, 10^3, 10^4$ and $M=10^5$ is plotted in Figs.~\ref{fig-y1} (a), (b), (c) and (d) respectively. For each block size, we show the variation of $P(y_1)$ for different values of the system size mentioned in each figure. 

As discussed earlier, to understand the effect of system size, we consider a multiplicative prefactor, $N^{\gamma_1}$ such that the quantity $N^{\gamma_1}P(y_1)$ collapses to a single function for a suitable value of $\gamma_1$. To determine $\gamma_1$, we consider the probability distribution at $y=y_c$ as a function of $N$. In  Fig.~\ref{Fig-sc} we plot the peak values of the probability distribution of normalized extreme activities $y_1$ as a function of $N$ for different block sizes. We observe that $P(y_1=y_c)\sim N^{-\gamma_1}$,  with $\gamma_1 = 1.25$. Note that this observation is consistent for different block sizes. Thus, with the observed value of $\gamma_1$, we plot the probability distribution rescaled namely $N^{\gamma_1}P(y_1)$ as a function of $y_1$ in the inset of Figs.~\ref{fig-y1} for different system sizes corresponding to each block size. We note that for each block size $M=10^2, 10^3, 10^4$ and $M=10^5$ the data collapse can be observed as shown in the inset of  Figs.~\ref{fig-y1} (a), (b), (c) and (d) respectively. For comparison, we plot the curve for the Gumbel family (black line)  (Eq.~\ref{Eq:Gumbel-PDF}) and the Weibull family (dashed red line). Note that the data collapse curve is in good agreement with the fitted curve of the Gumbel family for which the functional form is given by
\beq
f(y_1,\beta)=\frac{1}{\beta} \exp(-\exp(-y_1)-y_1).
\eqn

For comparison, we have also plotted the  Weibull family curve with $\xi=-0.19$ that appears to deviate from the observed data collapse. The fitted value of the data-collapse curve (cf. inset of Figs.~\ref{fig-y1}) suggests that the shape parameter of the GEV distribution is given by $\xi=0$. Thus, we infer that the probability distribution for the avalanche size converges into the Gumbel family (c.f. Eq.~\eqref{Eq:Gumbel-PDF} with $\xi=0$).

\begin{figure*}
\centering
    \includegraphics[width=160mm, height=130mm,angle=0]{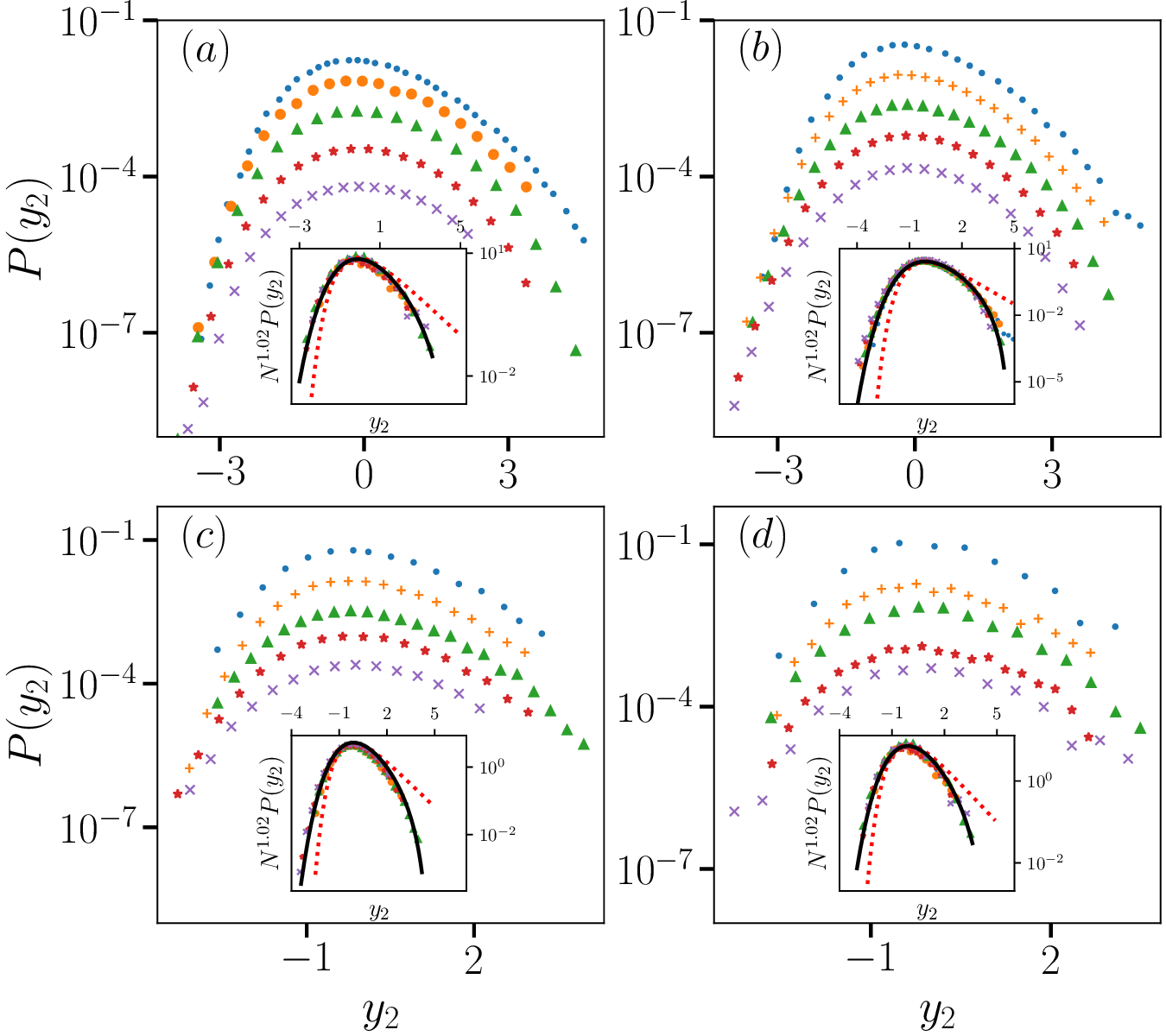} 
    \caption{Probability distribution of the GEV for extreme avalanche area corresponding to different values of the system size $L^{2i}$ using blue dots $ (i=3$), yellow points ($i=4$), green points ($i=5$), red points  ($i=6$), and purple points ($i=7$). These results for different block sizes are plotted (a) $10^2$, (b) $10^3$, (c) $10^4$, and (d) $10^5$. In the inset we plot the data collapse curve by plotting $N^{\gamma_2}P(y_2)$ versus $y_2$ for different block sizes using the data for various system sizes. The black curve in the inset is the plot for the Weibull  family and the red dashed line shows the curve for the  Gumbel family.}
    ~\label{fig-y2}
\end{figure*}

\begin{figure}[htb]
     \centering
     \includegraphics[width=60mm,height=50mm,angle=0]{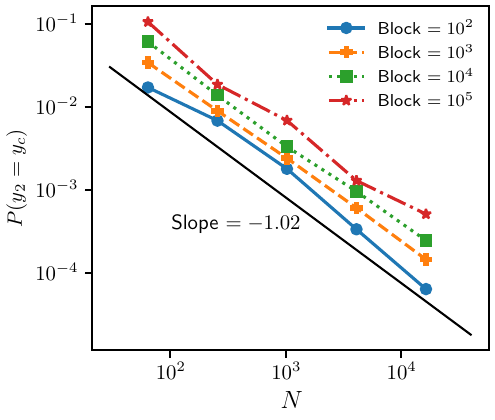}
     \caption{$P(y_c)~vs~N$ along with the fitted (black solid) curve $\sim N^{-1.02}$ for the variable $y_2$ that corresponds to extreme avalanche area.}
     \label{Fig-ac}
 \end{figure}

To study the behaviour of avalanche areas, we follow a similar prescription as outlined in the previous case.  The maxima in the avalanche area ($x_2$) can be normalized as
\beq
y_2=\frac{x_2-\la x_{2} \ra}{\sigma_2}
\eqn
where $\la x_{2} \ra$ and $\sigma_2$ are the mean and standard deviations for the variable $x_2$. The probability distribution of the rescaled variables $y_2$ follows a GEV distribution shown in Figs.~\ref{fig-y2} for different system sizes $N=L^2,~L=2^i$ where $i=3,4,5,6,7$. Figs.~\ref{fig-y2} (a), (b), (c) and (d) is plotted for different values of block sizes given by $M=10^2, 10^3, 10^4$ and $M=10^5$ respectively.  To understand the system size effect we consider a multiplicative prefactor $\gamma_2$ to observe the collapse on a single curve for different values of the system sizes. To obtain $\gamma_2$, we plot the behaviour of the peak values in the probability distribution of normalized extreme activities in Fig.~\ref{Fig-ac}. We observe that the probability distribution of the variable $y_2$ varies as  $P(y_2=y_c)\sim N^{-\gamma_2}$, with $\gamma_2 = 1.02$. With the help of the obtained value of $\gamma_2$ we show the data collapse of different system sizes in the inset of Figs.~\ref{fig-y2} (a), (b), (c) and (d) is plotted for different values of block sizes given by $M=10^2, 10^3, 10^4$ and $M=10^5$. In this case, the fitted value of the data-collapse curve (cf. Fig.~\ref{fig-y2}) suggests that the shape parameter of the GEV distribution is $\xi=-0.19$. Note that the data collapse is in good agreement with the rescaled probability distribution curve corresponding to the  Weibull family in the inset of Figs.~\ref{fig-y2} (cf. Eq.~\eqref{Eq:PDF-GEV}) with $\xi<0$. To highlight this effect, we plot, in the inset, the  Weibull family curve using black colour. For comparison, we also plot the Gumbel family curve using red dashed line.

\section{Summary}~\label{Sec:Conclusion}
 The SOC model (the 2d Abelian BTW sandpile) is one of the several dynamical models that may generate extreme events. We observe that the extreme avalanche activities show explicit system size dependence in the GEV probability distribution function.  We propose a scaling mechanism that provides a systematic approach to capture the FS scaling in GEV theory. This helps to distinguish the family of the GEV distributions. With finite system size, the avalanche activities in the BTW model follow a power law with upper cut-off~\cite{Yadav_2022}. This upper cut-off (or FS) makes the study interesting. We apply statistical physics and EVT and record statistics over a range to explore the universal scaling of the extreme magnitude of avalanche activities. The avalanche activities are parameterized by using the variables namely the avalanche size and the avalanche area. For the avalanche size, the shape parameter is observed to be $\xi=0$ indicating that the extreme avalanche activities converge to the Gumbel distribution. In the case of extreme avalanche area, we note that the rescaled distribution converges to the Weibull distribution with shape parameter value given by  $\xi=-0.19$.  Similarly, we also calculate the scaling exponent $\gamma_k$ that accounts for the effect of finite system size ($N^{-\gamma_k}$).  The scaling exponents for the extreme avalanche size and area were observed to be $\gamma_1=1.25$ and $\gamma_2=1.02$ respectively.  Thus the present study is helpful in exploring the system size dependence of the EVD associated with the extreme events. This provides a significant insight into one of the intriguing issues associated with extreme activities in SOC. Further, the proposed scaling analysis is such that the universal scaling function is found to be independent of the system size but with proper re-scaling of the variables. The rescaling of variables could be explained by the RG theory of extreme events~\cite{Calvo_2012, Gyorgyi_2010, Gyorgyi_2008}. Our findings could be extended to understand the nature of the SOC systems and could help us design schemes for dynamics governed by the SOC phenomena in natural systems.  We further investigated the evolution of extreme events with block size variations by analysing the statistical properties and observe that the evolution of extreme events in a system does not deviate substantially by changing the observation range. This scaling function has an intimate connection between extreme activities appearing on different length scales. Similar ideas may be implemented to weather modelling~\cite{Yao_2022}, integrable turbulence~\cite{Suret_2016, Kraych_2019}, to develop a surrogate model for characterizing extreme events. Whether or not such techniques will be applicable to other situations namely the geophysical phenomena or financial markets is still an open question.

\section{Acknowledgement}
AQ greatly acknowledges support from INSPIRE Fellowship (DST/INSPIRE Fellowship/IF180689), under the Department of Science and Technology, Government of India. We also thank Dr. Avinash Chand Yadav for useful suggestions.

\end{document}